\documentstyle[psfig]{europhys}

\def\And{{\rm and\ }}

\def\stars{\bigskip\centerline{***}\medskip}

\newif\ifboo \boofalse

\def\Review#1{\boofalse{\it #1},}
\def\Name#1{{\sc #1},}
\def\Vol#1{\ifboo Vol. {\bf #1}\else{\bf #1}\fi}
\def\Year#1{\ifboo #1\else(#1)\fi}
\def\Book#1{\bootrue{\it #1},}
\def\Page#1{\ifboo {\rm p. #1}\else{\rm #1}\fi}

\def\beq{\begin{equation}}
\def\eeq{\end{equation}}
\def\bea{\begin{eqnarray}}
\def\eea{\end{eqnarray}}

\def\tst{\textstyle}

\def\lam{\lambda}

\def\Dta{\Delta}

\def\hf{{1\over2}}
\def\tshf{\tst\hf}
\def\quar{{1\over 4}}
\def\tsqua{\tst\quar}

\def\xhat{\bf{\hat x}}

\def\zhat{\bf{\hat z}}

\def\ham{{\cal H}}
\def\bham{\bar\ham}
\def\ket#1{|#1\rangle}

\def\mel#1#2#3{\langle#1|#2|#3\rangle}

\def\bH{{\bf H}}
\def\bh{{\bf h}}
\def\bJ{{\bf J}}

\def\Fe8{Fe$_8$} 
\begin{document}
 
\euro{}{}{}{}
\Date{}
\shorttitle{ANUPAM GARG: DIABOLICAL POINTS IN THE Fe$_8$ SYSTEM}
 
\title{Diabolical points in the magnetic spectrum of Fe$_8$ molecules}
\author{Anupam Garg}
\institute{Department of Physics and Astronomy,
              Northwestern University\\
              Evanston, IL 60208, USA}
 
\rec{}{}
 
 
\maketitle
 
\begin{abstract}
The magnetic molecule Fe$_8$ has been predicted and observed to have a
rich pattern of degeneracies in its spectrum as an external magnetic field
is varied. These degeneracies have now been recognized to be
diabolical points. This paper analyzes the diabolicity and all essential
properties of this system using elementary perturbation theory.
A variety of arguments is gievn to suggest that an earlier semiclassical
result for a subset of these points may be exactly true for arbitrary
spin.
\end{abstract}
 
 
The molecular cluster [Fe$_8$O$_2$(OH)$_{12}$(tacn)$_6$]$^{8+}$
(or just \Fe8 for short) has
a total spin $J=10$ at low temperatures, and is described to a first
approximation by the spin Hamiltonian \cite{alb,sop,cam}
\beq
\ham  =  k_1 J_x^2 + k_2 J_y^2 - g\mu_B \bJ\cdot\bH, \label{ham}
\eeq
where $k_1 > k_2 > 0$, and $\bH$ is an external magnetic field. Thus
the axes $x$, $y$, and $z$ are hard, medium, and easy, respectively. EPR
measurements indicate $k_1 \approx 0.33$ K, $k_2 \approx 0.22$ K.

In the absence of any applied magnetic field, the spin of the molecule
has degenerate classical minima along the $\pm\zhat$ directions. Application
of a field cants the minima away from $\pm\zhat$, but the degeneracy is
preserved if $\bH$ is in the $x$-$y$ plane. This degeneracy is lifted
by quantum mechanical tunnelling between the low energy orientations.
It is of some interest to calculate the tunnel splitting $\Dta$, since
tunnelling plays an important role in the low temperature dynamics.
A few years ago, without knowledge of the relevance of Eq.~(\ref{ham})
to \Fe8, it was predicted \cite{agepl} that, for $\bH\|\xhat$, $\Dta$
would oscillate as a function of $H$, with {\it perfect} zeros at
certain values, and this effect was explained in terms of interference
arising from a Berry phase in the spin path integral. These
oscillations have now been seen by Wernsdorfer and Sessoli \cite{ws} using
a clever technique which enables Landau-Zener-St\"uckelberg (LZS)
transitions between the levels in question. The underlying value of $\Dta$
can be extracted from the observed LZS transition rate.

In addition to the predicted oscillations, however, Wernsdorfer and Sessoli
have also observed oscillations for certain non-zero values of $H_z$ as $H_x$
is swept. Villain and Fort \cite{vf} have noted that if $H_z$ is chosen properly,
these oscillations also represent perfect degeneracy, {\it i.e.}, $\Dta$ again
vanishes {\it exactly} at
isolated points in the $H_x$-$H_z$ plane [or the full three-dimensional 
$(H_x, H_y, H_z)$ space]. Thus, all the zeros of $\Dta$ are, in fact,
``diabolical points" in the magnetic field space. (This coinage is due to
Berry and Wilkinson \cite{bw}, as the shape of the energy surface when plotted
against two parameters in the Hamiltonian --- $H_x$ and $H_z$ in our case ---
is a double elliptic cone joined at the vertex, which resembles an
Italian toy called the {\it diavolo}.) Formulas for these points have been found by
Villain and Fort, and independently by the author \cite{aglt} (see below).

Diabolical points are of interest because of their rarity
in real-life physical systems. Indeed, the von Neumann-Wigner theorem
states that as a single parameter in a Hamiltonian is varied, an intersection
of two levels is infinitely unlikely, and that level repulsion is the rule.
It is useful to review the argument behind this theorem. Let the
energies of levels in question be $E_1$ and $E_2$, which we suppose to be far
from all other levels. Under an incremental perturbation $V$, the secular
matrix is
\beq
\left(
    \begin{array}{cc}
        E_1 + V_{11} & V_{12} \\
        V_{21}       & E_2 + V_{22}
    \end{array}
\right),    \label{sec}
\eeq
with $V_{21} = V^*_{12}$.
The difference between the eigenvalues of this matrix is given by
\beq
[(E_1- E_2 + V_{11} - V_{22})^2 + 4 |V_{12}|^2]^{1/2}, \label{edif}
\eeq
which vanishes only if
\beq
E_1 + V_{11} = E_2 + V_{22}, \quad\quad V_{12} = V^*_{12} = 0. \label{cond}
\eeq
Hence, for a general Hermitean matrix, three conditions must be satisfied for
a degeneracy, which in general requires at least three tunable parameters.
If the matrix is real and symmetic, the number of conditions and tunable
parameters is reduced to two \cite{va}.

An exception to this rule occurs when the Hamiltonian has some symmetry, when
levels transforming differently under this symmetry {\it can} intersect.
For the \Fe8 problem, the intersections when $\bH\|\xhat$ or $\bH\|\zhat$
can be understood in terms of symmetry \cite{ag3}, but those with both
$H_x$ and $H_z$ non-zero cannot.

The results reported in Ref. \cite{aglt} are based on a generalization
\cite{agprl}  of the discrete phase integral (or WKB) method \cite{dm,vs},
and are asymptotically accurate as $J\to\infty$. Villain and Fort use
an approximate version of the same method, with the additional
condition $k_1 - k_2 \ll k_1$. These calculations while involving
only elementary
methods of analysis, still entail the development of considerable
calculational machinery, and are quite long. Surprisingly, the full global
structure of the energy spectrum can be obtained by a much simpler
method---text-book perturbation theory in $k_2/k_1$ and the field components
$H_y$, $H_z$. This is an extension of an earlier calculation by
Weigert \cite{sw}, who analysed the problem for $H_y=H_z =0$.
In particular, one can rigorously
establish the existence of diabolical points, and find formulas for their
locations via a series of small calculations. It is
hoped that the simplicity of this approach will make the subject accessible
to a wide readership.

Before proceeding further, it is useful to
develop a scheme for labelling the eigenstates of $\ham$. Suppose first that
$\bH =0$, and $k_2 = k_1$. The states can then be labelled by the
eigenvalue $m$
of $J_z$, and the ground states are $m = \pm J$. If $k_2$ is now decreased,
states with $m$ differing by an even integer will mix. 
If $k_1 - k_2 \ll k_2$, or $J$ is large,
the barrier between $m=-J$ and $m=+J$
is large (see fig.~1), tunnelling is negligible, and we can find
states $\ket{m^*}$
which evolve continuously from $\ket m$, such that $\{\ket {m^*}\}$ are
eigenstates of $\ham$ to good approximation.
This approximation will continue to hold if the field $\bH$ is turned on, as
long as $|\bH| \ll H_c = 2k_1 J/g\mu_B$.
\begin{figure}
\centerline{\psfig{figure=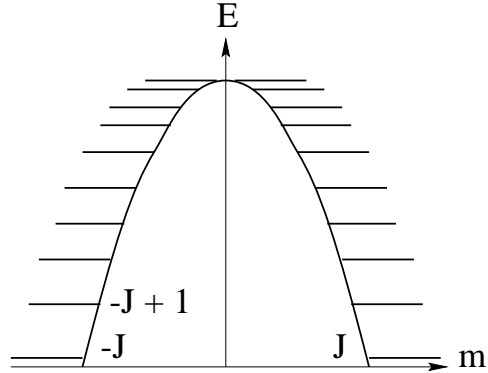}}
\caption{Schematic energy level diagram when $k_2 \approx k_1$.}
\label{fig1}
\end{figure}

The first set of diabolical points lies on the line $H_y = H_z =0$,
because $\ham$ is then invariant under a $180^{\circ}$ rotation about $\xhat$.
Levels with different parity under this operation can intersect as $H_x$ is
varied. In particular, the pseudo-ground states $m^* =\pm J$ are exactly
degenerate at a
sequence of $H_x$ values as found in Ref. \cite{agepl}. Since the symmetry is
destroyed if either $H_y \ne 0$ or $H_z \ne 0$,
so are the intersections, and the points are indeed diabolical. The same is
true of intersections of levels with $m^* = \pm(J - \ell)$, where $\ell$ is an
integer.

A similar argument applies when $\bH \| \zhat$, so another
set of diabolical points is expected
when $H_x = H_y =0$. In terms of fig.~1, the states which are degenerate are no
longer symmetrically located, and it is possible for say, $m^* = -J$ to be
degenerate with $m^* = J -1$. The new discovery by Wernsdorfer and Sessoli is
that the tunnel splitting between these states also oscillates as $H_x$ is now
varied. As mentioned above, these oscillations are also perfect, and the
corresponding diabolical points are not associated with any obvious symmetry
of $\ham$. (A similar situation holds in the spectrum of a particle confined
to a two dimensional triangular region \cite{bw}. Apart from an overall size,
which only affects the overall energy scale in a trivial manner, a triangle
is parametrized by two angles. Two sets of diabolical points
arise when the triangles are isoceles, but the rest
appear when the triangles are scalene with no special symmetry.)

We can thus classify the diabolical points by the $m^*$ numbers
of the levels which are degenerate. Let the state with predominantly negative
values of $m$ be labelled by $m^*_1$, and the other state by $m^*_2$. We define
$k = m^*_1 + J$, and $k' = m^*_2 -J$. In other words, counting from 0,
the $k$th level in the left well is degenerate with state number $k'$ in the
right well. When $k, k' \ll J$, the semiclassical analysis gives the location
of the diabolical point as ($H_y =0$ always)
\bea
h_x &=& {\sqrt{1-\lam} \over J}
                 \left[ J - \ell - {1\over 2}(k + k' +1) \right], 
    \label{Hxdia} \\
h_z &\approx&
       {\sqrt\lam \over 2J }(k-k'). \label{Hzdia}
\eea
Here, $\bh = \bH/H_c$ is a reduced field with $H_c = 2k_1J/g \mu_B$,
$\lam = k_2/k_1$, and $\ell$ is an integer.

Another way to label the degeneracies is to number the levels in order
of increasing energy, starting with 1 for the lowest level, and then
simply give the numbers of the two crossing levels.
Thus if the lowest two levels are degenerate ($k=k'=0$), we will
say that levels 1 and 2 cross, while for $k=0, k'=1$, or $k=1, k'=0$,
we would say that levels 2 and 3 cross. This labelling is not unique,
but we will find it convenient.

With this background, we now turn to our
calculations. Following Weigert \cite{sw} we regard the $k_2$ term in
Eq.~(1) as the perturbation, along with the $y$ and $z$ components of $\bH$.
It is convenient to divide all energies by $k_1$, and
write $\bham = \ham/k_1 = \bham_0 + \bham_1$, where
\bea
\bham_0 &=& J_x^2 - 2J h_x J_x, \label{H0} \\
\bham_1 &=& \lam J_y^2 - J (h_- J_+ + h_+ J_-), \label{H1}
\eea
where $J_{\pm} = J_y \pm i J_z$, $h_\pm = h_y \pm i h_z$. These notations
for $J_\pm$ are unconventional, but they are now convenient, as we will
take the quantization axis to be $x$, not $z$. 
We will label the eigenvalue of $J_x$ by $n$. To zeroth order, the energy
of state $n$ is given by
\beq
E_n^{(0)} = n^2 - 2 J h_x n, \label{en0}
\eeq
Levels $n$ and $n'$ are
approximately degenerate if $Jh_x = (n+n')/2$. To see if they are
exactly degenerate when $\bham_1$ is included, we find
the secular matrix $V$ to an appropriate order in perturbation theory,
and examine the conditions (\ref{cond}). We do this for a number of different
cases.

{\it Case 1 --- levels 1 and 2 cross.} Let the degenerate levels be $n_0$
and $n_0 + 1$, so that $Jh_x \approx (n_0 + \tshf)$.
For brevity, we label the states by A and B, and denote the
matrix elements $\mel {n_0 +1}{J_-}{n_0}$ etc. by $a_1$, $a_2$, $a_3$, etc.,
as indicated in fig.~2. Note that all $a_i$ can be chosen as real. To first
order in $\lam$ and $h_\pm$,
\bea
V_{AA} &=& \lam [J(J+1) - n_0^2]/2, \label{vaa} \\
V_{BB} &=& \lam [J(J+1) - (n_0 + 1)^2]/2, \label{vbb} \\
V_{AB} &=& - J h_+ a_2. \label{vab}
\eea
The conditions for diabolicity are thus
\beq
J h_x = (n_0 + \tshf)(1 - \tshf\lam),
        \quad h_y = h_z =0. \label{dia1}
\eeq
Writing $n_0 = J - \ell -1$, this is identical to Eqs.~(\ref{Hxdia})
and (\ref{Hzdia}) with $k=k'=0$, once we recognize that
$(1-\lam/2) = (1-\lam)^{1/2} + O(\lam^2)$. Since $-J \le n_0 \le J - 1$,
there are $2J$ such points.
\begin{figure}
\centerline{\psfig{figure=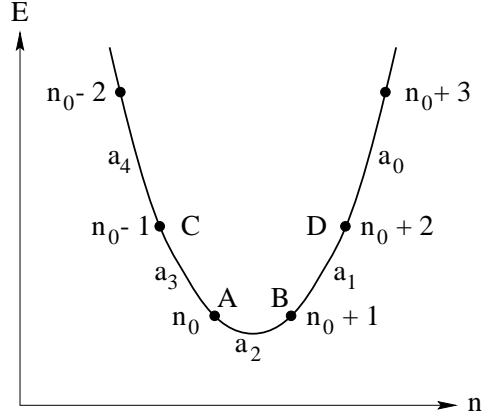}}
\caption{Energy level diagram for cases 1 and 3.}
\label{fig2}
\end{figure}

The conclusion that these points lie on the line $h_y = h_z =0$
is unchanged if we go to higher order. The relevant condition is clearly
that for off-diagonal elements. Contributions to the AB element
of the second order secular matrix arise from intermediate states
$n_0 +2$ and $n_0 -1$. A short calculation gives
$V^{(2)}_{AB} = \lam a_2 J (a_1^2 + a_3^2) h_-/8$. Adding this to
Eq.~(\ref{vab}) and setting the sum to zero, we again obtain the
conditions $h_y = h_z = 0$. 

{\it Case 2 --- levels 2 and 3 cross.} Let the lowest energy level
be $n_0$, and let $n_0 \pm 1$ be approximately degenerate. This
requires $J h_x \approx n_0$. Again,
we denote the states $n_0\pm 1$ by A and B, and the various matrix elements of
$J_\pm$ by $a_1$ to $a_4$ as in fig.~3. To $O(\lam)$, $V_{AA}$ and $V_{BB}$
are given by $\lam[J(J+1) - (n_0\pm 1)^2]/2$. The order $h_y^2$, $h_z^2$
contributions to the diagonal terms of the second order secular matrix are
found to both be equal to
$4 J^2(h_y^2 + h_z^2) [J(J+1) - n_0^2 +1]/3$. The interesting terms are
$V_{AB}$ and $V_{BA}$. Including first order pieces from $\lam J_y^2$,
and second order pieces from $h_{\pm}$, we get
\beq
V_{AB} = \left( \tsqua\lam  + J^2 h_+^2 \right)a_2 a_3. \label{vab2}
\eeq
For a diabolical point, therefore, the vector $\bh$ must have components
\beq
\bh = {1 \over J}\left[
        n_0 \bigl(1 - \tshf\lam \bigr), 0,
          \tshf{\sqrt\lam} \right]. \label{dia2}
\eeq
With $n_0 = J - \ell -1$, these are exactly the lowest order terms in
an expansion in $\lam$ of Eqs.~(\ref{Hxdia}) and (\ref{Hzdia}) with
$k=1$, $k'=0$. Since $-J + 1 \le n_0 \le J - 1$, there are $2J -1$
such points.
\begin{figure}
\centerline{\psfig{figure=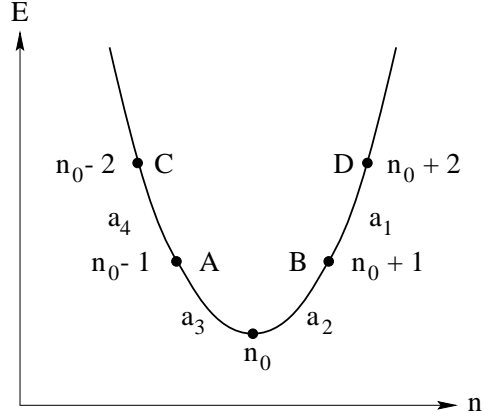}}
\caption{Energy level diagram for case 2.}
\label{fig3}
\end{figure}

{\it Case 3 --- levels 3 and 4 cross.} This case can arise either with
$k=k'=1$, or with $k=2$, $k'=0$, but we shall be able to distinguish
between these. Refering to fig.~2 again, the degenerate levels are
$n_0 - 1$ (C) and $n_0 + 2$ (D). Equality of $E_C$ and $E_D$ again
requires $Jh_x \approx (n_0 + \tshf)$. To first order in $\lam$,
$V_{CC} - V_{DD} = -3 \lam (n_0 + \hf)$, so that the diagonal elements
are equal when
\beq
Jh_x = (n_0 + \tshf)(1 - \tshf\lam). \label{dia3x}
\eeq
As in case 2, it is the off-diagonal
term which is of greater interest. The secular matrix is now diagonal in first
order, and off-diagonal terms only arise in second and higher orders.
Second order terms arise from the combination of one $h_\pm J_\mp$ term
and one $\lam J_y^2$ term, while third order terms arise from three
$h_\pm J_\mp$ terms. The net result is
\beq
V_{CD} = - \tsqua(h_+^2 J^2 + \lam) h_+ J a_1 a_2 a_3. \label{vcd}
\eeq
This can vanish in two ways. The first is to have $h_y = h_z =0$, 
in which case the diabolical field is given by Eq.~(\ref{dia1}) again.
This case corresponds to $k = k' = 1$.

The second way for $V_{CD}$ to vanish is for the factor in parentheses
in Eq.~(\ref{vcd}) to vanish. This happens when
\beq
h_y = 0, \quad  h_z = {\sqrt\lam}/J. \label{dia3yz}
\eeq
In conjunction with Eq.~(\ref{dia3x}), this is seen to be the
same as Eqs.~(\ref{Hxdia}) and (\ref{Hzdia}) with $k=2$, $k'=0$,
and $n_0 = J - \ell -2$.

It is clear that this procedure gets rapidly more tedious if we
apply it to cases with larger $k$ and $k'$. It is more useful to
consider higher order perturbative corrections for the cases treated
above. In the argument leading to Eq.~(\ref{dia3x}), {\it e.g.},
we have only gone up to $O(\lam)$.  It is obvious that inclusion
of higher order terms can at best alter the value of $h_x$ at the
diabolical point by terms of order $\lam^2$, $h_y^2$, and $h_z^2$,
but cannot destroy the existence of a perfect degeneracy. The same
argument applies in all the other cases, and constitutes a constructive
proof of the existence of diabolical points.

It is particularly interesting to investigate the subset of diabolical
points on the line $H_y = H_z = 0$ in greater depth.
As noted before, these points
correspond to $k=k'$, and the degenerate levels have $n$ quantum
numbers differing by an odd integer. Thus they can never be coupled by
the remaining perturbation $\bham_1 = \lam J_y^2$, and the problem is
effectively one of non-degenerate perturbation theory. 
Let us consider case 1 first. The second order correction to the energy
of state A arises from the intermediate states $n_0 \pm 2$, and to that
of state B from $n_0 - 1$ and $n_0 +3$. It suffices to find the energy
denominators assuming that $Jh_x = (n_0 + \hf)$. A short calculation
gives
\beq
\left( \begin{array}{c}
V^{(2)}_{AA} \\ V^{(2)}_{BB} \end{array} \right)
 =
-{2 \over 3} \lam^2 \left[
\bigl[ J(J+1) - (n_0^2 + n_0 + 1) \bigr]^2
 + 2n_0^2 +
 \left( \begin{array}{c}  -n_0 -1 \\ 5n_0 + 2 \end{array} \right)
                          \right].
\label{v2aa}
\eeq
Along with Eqs.~(\ref{en0})-(\ref{vbb}), this means that to
$O(\lam^2)$, the states are degenerate when
\beq
J h_x = (n_0 + \tshf)(1 - \tshf \lam -{\tst{1 \over 8}}\lam^2),
   \label{dia12}
\eeq
which is precisely what Eq.~(\ref{Hxdia}) also gives.

In the same way, for the subcase $k=k'=1$ of case 3, we obtain
\beq
\left( \begin{array}{c}
V^{(2)}_{CC} \\ V^{(2)}_{DD} \end{array} \right)
 =
-{1 \over 40} \lam^2 \left[
\bigl[ J(J+1) - (n_0^2 + n_0 - 1) \bigr]^2
 - 6 n_0^2 +
 \left( \begin{array}{c}  9 n_0 - 4 \\ -21 n_0 -19 \end{array} \right)
                          \right].
\label{v2cc}
\eeq
Including lower order terms, the condition for degeneracy
is found to be identical to Eq.~(\ref{dia12}).

The fact that the two pairs of states $k=k'=0$, and $k=k'=1$ are
simultaneously degenerate (at least to order $\lam^2$), is very
striking. Calculations to $O(\lam^2)$ were in fact done by Weigert
\cite{sw}, but he did not perform them sufficiently explicitly, and
reached the opposite conclusion, {\it i.e.}, that the degeneracy
conditions would be different. It is clear, however, that this
equality is a result of the simple form of $\ham$, and is violated when
higher anisotropies such as $(J_x \pm i J_y)^4$ are included.

The second striking feature about the result (\ref{dia12}) is that there
are no terms like $\lam^2J^2$ or $\lam^2n_0^4$ etc. on the right hand side,
and that it is agrees precisely with the semiclassical answer. Since the
latter is obtained in a very different limit, namely, $J \to \infty$,
it begins to raise the suspicion that it might be exact. To test this
suspicion, we have carried the calculation for case 1 to order $\lam^3$.
For this, not only must we find $V^{(3)}_{AA}$ and $V^{(3)}_{BB}$, but we
must also keep $O(\lam)$ corrections in the energy denominators in the
calculations for $V^{(2)}_{AA}$ and $V^{(2)}_{BB}$, since $Jh_x$ depends
on $\lam$ at the diabolical point.  The resulting calculation is
lengthy, but is efficiently done using MAPLE. Almost miraculously, all 
powers of $J$ multiplying $\lam^3$ cancel, as do terms $\lam^3 n_0^j$
with $j \ge 2$, and the contribution to $E_A - E_B$ is just
$\lam^3 (2n_0 + 1)/16$. The condition for degeneracy thus becomes
\beq
J h_x = (n_0 + \tshf)(1 - \tshf \lam -{\tst{1 \over 8}}\lam^2
   - {\tst{1 \over 16}}\lam^3).
   \label{dia13}
\eeq
It will not have escaped the reader that the last factor equals
$(1-\lam)^{1/2}$ to $O(\lam^3)$!

It is useful to consider the structure of the perturbation series to
higher order in $\lam$. It is clear that we cannot get
negative powers of $J$ in the formula for $Jh_x$; instead it generates
positive powers. Although the low order analysis suggests otherwise,
in principle we should expect terms such as $\lam^N J^K(J+1)^K$ with
$0<K< N-1$ in $N$th order.  Such terms would be reminiscent of an
asymptotic series, and would
signal a zero radius of convergence. Such a situation would be very
odd in our problem since the perturbation $\lam J_y^2$ does not appear
to be singular. Although plausible, this is far from a complete argument
that such terms are in fact absent, since
we have not excluded terms such as $\lam^N n_0^{N-1}$ in $N$th order.

Further evidence that the result (\ref{Hxdia}) is exact comes from
looking at low values of $J$. We have done this for $J$ up to 2.
For $J=1/2$, there is nothing to prove
as the only degeneracy is at $h_x = 0$, which is also guaranteed by
Kramers's theorem. For $J=1$, the energies are directly
found to be
$E_{\pm 1} = 1 +\hf\lam \mp (4 h_x^2 J^2 +{\quar}\lam^2)^{1/2}$,
and $E_0 = \lam$, so $E_1 = E_0$ when $J h_x = (1-\lam)^{1/2}/2$.
For $J=3/2$, $\bham$ separates into two $2 \times 2$
matrices in the $J_x$ basis, which we call $M_1$ and $M_2$. Both
eigenvalues of $M_1$ coincide with those of $M_2$ at
$h_x =0$. This is again Kramers's degeneracy. In addition, one eigenvalue
of $M_1$ coincides with one of $M_2$ precisely when
$h_x = 2(1-\lam)^{1/2}/3$. For $J=2$, $\bham$ separates into a
$3\times 3$ matrix ($M_1$) and a $2\times 2$ matrix ($M_2$). The expected
degeneracies are at $h_x = (1-\lam)^{1/2}/4$ and $3(1-\lam)^{1/2}/4$.
At the second value of $h_x$, one eigenvalue of $M_1$ indeed coincides
with one of $M_2$. At $h_x = (1-\lam)^{1/2}/4$, however, {\it two
distinct} $M_1$ eigenvalues coincide with two $M_2$ eigenvalues.
Thus we again see the simultaneous degeneracy of two sets of levels
($k=k'=0$, and $k=k'=1$), leading us to believe that this feature is
also generally true. A rigorous proof of these conjectures remains
an open problem.

%
%
\stars
This work is supported by the National Science Foundation through
Grant No. DMR-9616749.
%
%

\end{document}